\documentclass[conference]{IEEEtran}

\usepackage{color}
 \usepackage{url}
 \usepackage{psfrag}
\usepackage{cite}
\usepackage{hyperref} 
\usepackage{subfig}\usepackage[T1]{fontenc}
\usepackage[utf8]{inputenc}
\usepackage{floatrow}
\usepackage{algorithm,algpseudocode}
\usepackage{graphicx}

\makeatletter

\newcommand{\Rmnum}[1]{\expandafter\@slowromancap\romannumeral #1@}
\makeatother

\begin{document}
%
\title{Efficient Decoding of Synchronized Colliding LoRa Signals}

\author{%
\textsc{Samira Abboud, Nancy El Rachkidy, Alexandre Guitton} \\
\normalsize Universit\'e Clermont Auvergne, CNRS, LIMOS, F-63000 Clermont-Ferrand, France\\ 
\normalsize {samira.abboud@isima.fr, nancy.el\_rachkidy@uca.fr, alexandre.guitton@uca.fr}}

\maketitle

\begin{abstract}
In LoRa (Long Range), when a collision occurs in the network, each end-device has to retransmit its colliding frame, which reduces the throughput, and increases the energy consumption of the end-devices and the delay of the frames. In this paper, we propose an algorithm to decode colliding synchronized LoRa signals and thus improve the overall performance of the network. Indeed, we use successive transmissions of bitmaps by the end-devices to determine the correct symbols of each colliding frame, instead of retransmitting the whole frames. Simulation results show that our algorithm is able to significantly improve the overall throughput of LoRaWAN, and to decrease the energy consumption and the delay of the transmitters.
\end{abstract}

 \textit{keywords} - 
LoRa, LoRaWAN, collision cancellation, synchronized signals.
 
 \IEEEpeerreviewmaketitle

\section{Introduction}
Internet of Things (IoT) installations are becoming a reality  and  networks  are  being  deployed  to realize smart cities, such as transportation and vehicular traffic, healthcare, smart parking lot, surveillance systems and environmental monitoring applications \cite{talari2017review}. For example, the climate conditions of the greenhouse or agricultural field, such as humidity, temperature, fires, earthquake, flooding, volcanoes, air pressure, water level, and soil moisture are monitored, where sensors may send data to the server over non-saturated network. Many of these IoT installations rely on Low-Power Wide-Area  Network  (LPWAN) technologies. These emerging technologies such as Long Range (LoRa)~\cite{LoRaSemtech}, Sigfox~\cite{SigFox}, RPMA~\cite{IngenuRPMA} and Weightless~\cite{Weightless} enable power-efficient wireless communication over very long distances.

LoRa~\cite{LoRaSemtech} is a widely deployed LPWAN technology and is considered by a large number of industries as a base for their IoT applications. LoRa  uses  orthogonal  transmission  settings (such as frequency, spreading factor) to reduce collisions. However, collisions cannot be totally avoided even when considering the capture effect. Current LoRa deployments use a default behaviour for retransmissions of colliding frames where each transmitter (e.g. the end-device) has to retransmit the whole colliding frame, which leads to reduce the overall throughput, and to increase the energy consumption and the delay. For these reasons, it is desirable to find a method for decoding the colliding signals, while decreasing the delay and the energy consumption of the transmitters, and improving the throughput. 

In this paper, we consider non-saturated traffic conditions where the performance of both LoRaWAN and our proposed algorithm are analyzed in terms of the actual observed throughput, delay and energy consumption. 
Besides, we consider the case of fully synchronized signals. 

We show that overlapped symbols can be extracted by the receiver, although it is not possible to determine to which frame each symbol belongs. We propose an algorithm that aims to decode such superposed signals and to reduce the impact of collisions. The proposed algorithm relies on sending bitmaps by the end-devices in order to determine the correct symbols of the frames, instead of retransmitting the whole frames. This algorithm always succeeds to retrieve the frames from the superposed signals regardless of the number of colliding transmitters. Furthermore, the delay, the energy consumption, and the throughput are improved compared to LoRaWAN.
Simulation results confirm the effectiveness of our proposed algorithm.

The rest of the paper is organized as follows. Section II presents a description of LoRa and LoRaWAN and some related works. Section III describes our proposed algorithm used for decoding fully synchronized signals. Section IV presents our simulation results. Finally, Sect. V concludes this paper.

 \section{State of the art}
 In the following, we first describe the physical layer LoRa and the media access control (MAC) protocol LoRaWAN. Then we present some of the related works.

\subsection{The LoRa physical layer}
 Long  Range  (LoRa)  is  a  proprietary  Chirp Spread  Spectrum  (CSS) modulation  technique  by  Semtech \cite{LoRaSemtech}. LoRa main parameter is the Spreading Factor (SF) which has an influence on the transmission duration, the energy consumption, the robustness and the communication range. SF defines the number of bits encoded into each symbol, and can vary between 7 and 12. Each symbol encodes one of $2^{SF}$ values, which cover the entire frequency band. When the maximum frequency of the band is reached, the frequency wraps around, and the increase in frequency starts again from the minimum frequency.
A high spreading factor increases the Signal to Noise Ratio (SNR) and therefore the receiver sensitivity and the range of the signal. However, it reduces the transmission rate and thus increases the transmission duration and the energy consumption. The SFs in LoRa are usually considered orthogonal. Consequently, concurrent transmissions with different SF do not interfere with each other, and can be successfully decoded. 

\subsection{The LoRaWAN protocol}
LoRaWAN \cite{alliance2015lorawan} is an open-standard protocol which defines the MAC layer for LPWAN technology. LoRaWAN is designed by the LoRa Alliance \cite{loraAlliance} and operates on top of LoRa. LoRaWAN architecture is composed from end-devices that are connected to the network server through gateways which relay messages. LoRaWAN enables three classes of operation for end-devices: class A, class B, and class C. In class A, end-devices send data when ready and wait for an acknowledgment from the network server. Then, they switch to sleep mode to save energy until the next transmission. The delay between two transmissions has to be larger than 99 times the duration of the frame transmission in order to respect the duty cycle of 1\%. In class B, which is optional, end-devices have additional scheduled receive periods to allow downlink communications with a bounded delay. In class C, which is also optional, end-devices are always active. 
For Europe region, the bandwidth of the channel is equal to 125 kHz for data rate 0 to data rate 5, and 250 kHz for data rate 6. SF7 is used for data rate 5 and data rate 6, and SF12 is used for data rate 0.

\subsection{Related work on LoRa}
Gateways in LoRa are able to decode superposed signals when they are sent on different SFs or on different channels. When signals are sent on the same channel and with the same SF, they risk to collide, unless the strongest signal is captured by the receiver. Since LoRa is a new technology, the amount of research on different aspects of both LoRa technology and LoRaWAN networks is limited. There have been a couple of works dealing with LoRaWAN collisions.
Some researchers such as \cite{rahmadhani2018lorawan}, \cite{ferre2017collision},\cite{reynders2016chirp} and\cite{petajajarvi2017performance} have studied the collisions in LoRa and their impact on the throughput.
 
In \cite{rahmadhani2018lorawan}, the authors presented an in-depth investigation of LoRaWAN frame collisions and the capture effect in particular through various experiments. They focused on correct reception of data at the application, instead of at the gateway, and they consider multi-gateways, and dense scenarios to obtain insight into collisions within actual networks, in order to investigate under which circumstances collisions lead to frame loss. For example, their experiments showed that using multiple gateways instead of a single gateway increases the probability of receiving correct frames. Furthermore, they found that most frames hardly reached the more distant gateways, which is possibly due to the low spreading factor used most of the time. Collisions can also aggravate the situation, especially for the frames that use high SFs and required longer time on air.

In \cite{ferre2017collision}, the author proposed an analysis of packet collision and packet loss probabilities in LoRaWAN, and developed  theoretical  expressions  for  both  of them. The author showed that his theoretical  expressions  are  more  accurate  than  a  Poisson distributed process to describe the collisions.

In \cite{reynders2016chirp}, the authors made a study regarding the CSS modulation technique. They show that some CSS symbols are not orthogonal. Their simulations show that the achievable range of the CSS technique is lower than an ultra-narrowband solution, but the robustness against interference is higher.

In \cite{petajajarvi2017performance}, the authors provide an analysis and report experimental validation of the various performance metrics of the LoRa technology. The LoRa modulation is based on CSS, which enables low-quality oscillators in the end-device, and faster and more reliable synchronization. Therefore, LoRa seems to be a promising option for implementing communication in many diverse IoT applications. Authors first  overviewed the features and analyzed the scalability of LoRa network. Then, they introduced setups of the performance measurements. Their results showed that using the transmit power of 14 dBm and the highest spreading factor of 12, more than 60\% of the packets are received from the distance of 30 km on water. With the same configuration, they measured the performance of LoRa communication in mobile scenarios. Their results revealed that at around 40 km/h, the performance gets worse, because duration of the LoRa modulated symbol exceeds coherence time. 

\subsection{Related work on synchronized signals on LoRa}
Few works have been done to study the collisions among synchronized signals in LoRa as in the following papers \cite{liao2017multi}, \cite{eletreby2017empowering} and \cite{rachkidy2018decoding}.

In \cite{liao2017multi}, the authors worked on constructing an efficient multi-hop network based on the sub-GHz LPWAN technology. They investigated the combination of LoRa and concurrent transmission (CT) which is a flooding protocol that considers synchronized packet collisions that happen when multiple relays perform immediate retransmissions at the same time. They found that, due to the time domain and frequency domain energy spreading effect, LoRa is robust to the packet collisions resulting from CT. They found the receiver performance under CT can be further improved by introducing timing offsets between the relaying packets. Therefore, they proposed a timing delay insertion method, the offset-CT method, that adds random timing delay before the packets while preventing the timing offset from diverging over the multi-hop network. Their experiments demonstrate the feasibility of CT-LoRa multi-hop network, and the performance improvement brought by the CT method. Their results showed that CT-LoRa  experiences a high packet reception rate performance under the typical multiple-building area networks scenario. 
Moreover, they showed that LoRa survive the CT purely by capture effect which is considered in order to increase the probability of decoding colliding LoRa signals. However, if after taking into account the capture effect, the colliding signals are not decoded, they are considrered lost. In this paper, we decode LoRa colliding signals without considering the capture effect (although with the capture effect  collisions are less encountered).

In \cite{eletreby2017empowering}, the authors have presented Choir which is a system that improves throughput and range of LPWANs in urban environments. Choir proposed a novel approach that exploits the natural hardware offsets between low-power nodes to separate collisions from several LPWAN transmitters using a single-antenna LPWAN base station. Further, Choir allows groups of LoRaWAN sensor nodes with correlated data to reach the base station, despite being individually beyond communication range. Choir directly improves the throughput of dense urban LPWANs by decoding transmissions from multiple nodes simultaneously with minimal coordination overhead. Choir recognizes that in practice, two signals who are synchronized in time can be separated by exploiting the natural hardware imperfections of the two radios. Specifically, signals from the two transmitters are likely to experience a small frequency offset, due to a difference in the frequency of their oscillators. This would result in the two chirps being slightly offset in frequency. Here are the differences between \cite{eletreby2017empowering} and this paper: 1) While Choir allows collisions from multiple transmitters to be decoded, its gains are bounded and limited when increasing the number of nodes, and the possibility of overlapping frequency offsets that increases with collisions from a larger number of transmitters. In this paper, we can manage a large number of transmitters by sending small bitmaps to save energy and further increase LoRa throughput. 2) If Choir fails to decode the synchronized colliding frames, it must retransmit the entire frames as in the traditional LoRaWAN protocol. Furthermore, if the collision continues to reproduce again between the same transmitters, it remains indecodable because the transmitters do not change their frequencies. In this paper, we rely on sending bitmaps and not to retransmit the entire frames. 3) Choir relies on the frequency offset to separate and decode synchronized interfering transmissions. In this paper, we did not consider frequencies. Consequently, we can combine both algorithms: Choir in first, and if it fails to decode colliding signals, our proposed algorithm comes in second to decode colliding signals.

In \cite{rachkidy2018decoding}, the authors have proposed two algorithms to decode some cases of collisions of LoRa signals. The first algorithm is used when superposed signals are slightly desynchronized, and the second algorithm is used when superposed signals are completely synchronized. Authors observe that the first algorithm is able to significantly increase the throughput, by decoding many collisions of two signals, and some collisions of three signals. On the other side, the second algorithm has improved the throughput, by decoding both signals when exactly two signals are colliding. The second algorithm requests any of the two colliding transmitters to retransmit its frame. Hence, when one frame is retransmitted, the algorithm is able to decode it, and to deduce the other colliding frame by elimination. The authors have considered the case of only two synchronized signals.
In this paper, we are improving this by reducing the amount of data retransmitted, and by considering two or more synchronized signals.

 \section{Propositions}
 In this section, we present our contributions for the physical and MAC layers. We propose an algorithm to decode superposed signals that are fully synchronized. These signals are received on the same channel, with the same SF, and with the same received signal power. The proposed algorithm can not be applied on LoRaWAN protocol since most communications in LoRaWAN are desynchronized. Therefore, we introduce a new MAC layer which could be used on top of LoRa.
 
\subsection{Physical algorithm}
\begin{figure}[h]
	\centering
	\includegraphics[width=60mm,scale=0.5]{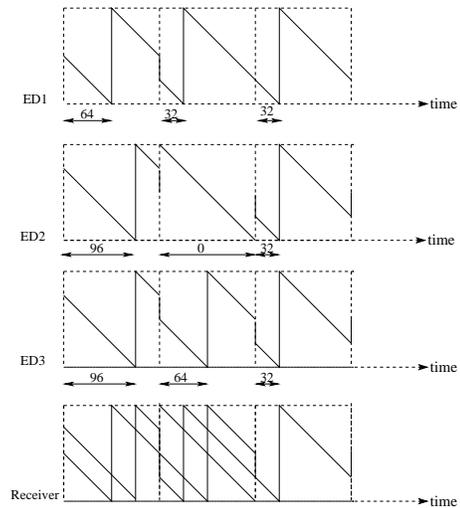}
	\caption{Superposition of three synchronozed signals.}
	\label{FigSuperposition}
\end{figure}

In this subsection, we consider the superposition of signals from transmitters that are fully synchronized as in \cite{rachkidy2018decoding}.  Recall that papers \cite{liao2017multi} and \cite{eletreby2017empowering} have ensured the feasibility of the synchronization among LoRa signals, by implementing a real LoRa system where transmissions were synchronized in time. Figure~\ref{FigSuperposition} shows an example of the reception of three fully synchronized signals under SF7. The figure shows that the signals start at the same time. Frame transmitted by end-device 1 (referred to ED1) is $f_{1} = (64, 32, 32)$, frame transmitted by ED2 is $f_{2} = (96, 0, 32)$, and frame transmitted by ED3 is $f_{3} = (96, 64, 32)$. In this figure, we can see that the receiver can extract symbols \{64, 96\} during the first symbol duration, symbols \{0, 32, 64\} during the second symbol duration, and symbol \{32\} during the third symbol duration.

However, the receiver is not able to determine to which frame each symbol belongs.

  \begin{figure*}[h]
 \psfrag{b11}{$0$}
  \includegraphics[width=150mm,scale=0.8]{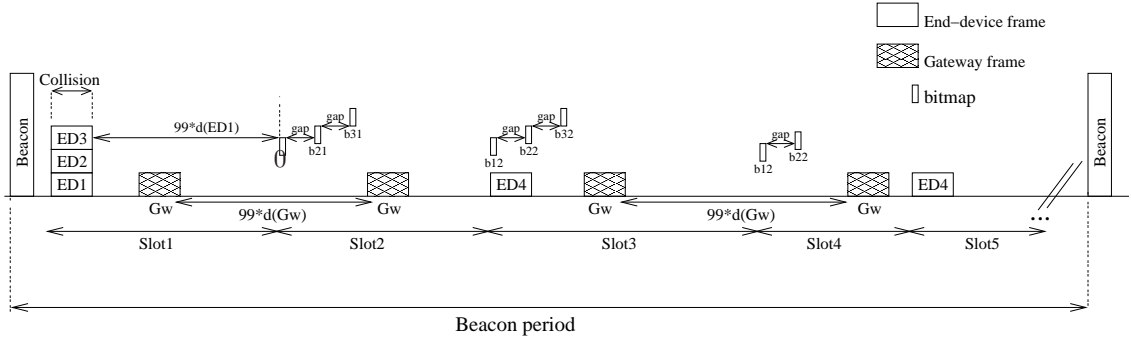}
  \caption{New MAC protocol depicted for the superposition of three synchronized signals.}
   \vspace{10pt}
  \label{FigMacBeaconSynch}
\end{figure*}

\subsection{Our MAC protocol}

In this subsection, we present and explain our MAC protocol used to decode synchronized colliding signals.

\subsubsection{\underline{Description of MAC protocol and timing computation}}
Figure~\ref{FigMacBeaconSynch} shows a MAC protocol implementing our proposed algorithm depicted for four end-devices. Beacons are sent by the gateway to synchronize the communications. We assume that the time is divided into slots and that transmissions on the same slot are fully synchronized. When a collision occurs between frames, the gateway stores all superposed symbols at each symbol duration. Then, it sends a frame built from these symbols. The gateway waits for bitmaps from the end-devices in order to decode the colliding frames, where each bit corresponds to the symbol chosen in the gateway frame. The gateway frame contains, in addition to the symbols, the order of the end-devices using an identifier on one symbol. As long as the frames sent by the end-devices are not yet decoded, the gateway sends a new frame, and the end-devices reply by sending new bitmaps.
  
To further describe our new proposed MAC protocol, we develop a timing computation model for the transmission process. Specifically, we consider separately the first and the subsequent transmission attempts.

 \textit{The first frames transmission:}
The first transmission attempt of the end-devices frames is made on Slot1, where three end-devices ED1, ED2 and ED3 are initially fully synchronized, and sent their uplink frames at the same initial start time $t_{0}$. This causes a collision at the gateway Gw which stores the superposed symbols, and sends a frame composed from these superposed symbols (See Section \ref{subsection:GuessingFrame} for an example of the frame sent by the gateway).
\\
The start time of the first gateway frame ($t_{0_{Gw}}$) is equal to the initial start time of the EDs frames ($t_{0}$) plus the duration $d_{ED}$ (i.e time on air) of the EDs frames as follows:
$t_{0_{Gw}} = t_{0} + d_{ED}$

\textit{The bitmaps first transmission:}
Each bitmap of a given end-device is separated from the bitmap of the previous and the next end-devices by a small amount of time called guard interval (i.e $gap$). This guard time ensures that the bitmap of an end-device does not collide with the bitmap of another end-device, even when considering clock drifts. 
\\
By referring to Fig.~\ref{FigMacBeaconSynch}, each end-device in Slot2 sends a bitmap in reply to the Gw frame. The bitmaps ${b_1^1}$, ${b_1^2}$, and ${b_1^3}$ are sent by ED1, ED2 and ED3 respectively, after 99 times the duration of the frame transmission of ED1, ED2 and ED3. In addition, for each end-device $x$, its bitmap $b_i^x$ is delayed to avoid a collision with $b_i^{(x-1)}$ and $b_i^{(x+1)}$. In other words, for an end-device $x$, the start time $t_{b_i^x}$ of its bitmap number $i$ should respect the following rule:
$(t_{b_i^{(x-1)}} + d_{b_i^{(x-1)}})\times(1+\Delta_{max}) \leq (t_{b_i^x})\times(1-\Delta_{max})
$ where $d_{b_i^{(x-1)}}$ is the time on air of the the previous bitmap, and $\Delta_{max}$ is the maximum drift.
\\
Hence, the start time $t_{b_i^x}$ of a bitmap $b_i^x$ is given by the following equation:
$
t_{b_i^x} = t_{b_i^{(x-1)}} + d_{b_i^{(x-1)}} + gap
$  with $ 
t_{b_i^{(x-1)}} = t_{Slot} + (x-2)\times d_{b_i^{(x-1)}} + (x-2) \times gap
$
where $t_{Slot}$ is the start time of the current slot.
\\
Moreover, the duration of a slot is given by the following:
$  \begin{array}{l}
d_{Slot} = \max (d_{ED} + d_{Gw}, d_{b_i^x} \times x+gap \times (x-1) + d_{Gw})
  \end{array}
$ where $d_{ED}$ is the duration of the end-device frame, and $d_{Gw}$ is the duration of the gateway frame.
 \\
In relation to the slot duration, we set the start time of the last slot as follows:
$ t_{Slot_{nmaxslots}} = d_{Slot} \times (nmaxslots-1)
$ where $nmaxslots$ is the maximum number of slots.  
 
\textit{The bitmaps subsequent transmissions:} The start time of a bitmap for the subsequent tranmissions is given by $\begin{array}{l}
 t_{b_i^x}  = \max (t_{b_{(i-1)}^{x}} + 100 \times d_{b_{(i-1)}^{x}} , t_{Gw} + 100 \times d_{Gw})
 \end{array}
$ where the end-device should respect the duty cycle regulation, and should wait for the gateway frame before sending its bitmap $b_i^x$ with $i > 1$. 

\textit{Specific case:}
The general MAC algorithm contains slots where collisions between bitmaps and the frames of end-devices may occur. As shown in Slot3, the frame of ED4 collides with the bitmaps ${b_2^1}$ and ${b_2^2}$ related to ED1 and ED2 respectively. The gateway receives the bitmap ${b_2^3}$ of ED3 and decodes it, while bitmaps ${b_2^1}$ and ${b_2^2}$ are not received. This leads ED1 and ED2 to retransmit their colliding bitmaps in Slot4. Also, ED4 (which corresponds to $x=4$) retransmits its colliding frame as shown in Slot5 while respecting the duty cycle of 1\% as follows:
$ t_{ED_{x}} = t_{ED_{x-1}} + 100 \times d_{ED}
$.
In addition, the frame of a given end-device may collide with all the bitmaps of the other $y$ end-devices that are sent on the same slot if: 
$ d_{ED} \geq y \times d_{b_i^y} + (y-1) \times gap
$
 \subsubsection{\underline{Description of our algorithm}}
 Our algorithm described by Algorithm \ref{alg:decoding} is used to decode fully synchronized colliding signals for \textit{x} transmitters (i.e end-devices), with $\textit{x} \geq 2$. 
\\ In this paragraph, we explain how the receiver detects and decodes the colliding frames sent by $x$ colliding end-devices. Indeed, when a collision occurs on the $x$ EDs, the gateway can not decode the colliding frames, and is not able to determine to which frame each symbol belongs. Hence, the gateway considers that all the end-devices frames contain missing symbols, represented by * in Algorithm \ref{alg:decoding}. Meantime, the gateway sends a frame built from the superposed symbols. At this step, each ED $x$ replies to the gateway frame by sending a bitmap $b_{i}^{x}$ (which corresponds to the bitmap number $i$). For each $bit$ at position $j$ in $b_{i}^{x}$, the algorithm checks the value of the $bit$. If $bit$ is equal $1$, then the algorithm replaces * in the frame of ED $x$ with the current symbol $j$ of the gateway frame. On the other hand, if the $bit$ is equal $0$, the algorithm checks if the number of superposed symbols at the current position $j$ is equal to $2$. If it is the case, then the algorithm replaces the * with the other current symbol at position $j$ of the superposed symbols. In addition, the algorithm verifies if at position $j$, all the symbols of the EDs have been decoded (i.e not equal to *), and if there is still a missing symbol (i.e *) in a frame of another ED $y$. If it is the case, then the algorithm replaces the * in the symbol $j$ of $y$ by the remaining current symbol in the same position $j$ of the superposed symbols. As long as there are missing symbols that can not be decoded by the gateway, the process is repeated until the decoding of all colliding signals.
It is worth mentioning that this algorithm runs in polynomial time.
\begin{algorithm}
\caption{Decoding of fully synchronized superposed signals.}
\begin{minipage}{0.98\linewidth}
\label{alg:decoding}
\begin{algorithmic}[1]
\While {a frame contains *}
\State the gateway sends a frame
   \For{each end-device $x$ }
    \State $x$ sends a bitmap $b_{i}^{x}$
        \For{each $bit$ at position $ j$ in $b_{i}^{x}$ }
            \If{$bit = 1$}
                \State symbol $j$ of the frame $f_x$ $\leftarrow$  symbol $j$ \State of the gateway frame
            \ElsIf{$bit = 0$ and the number of \State superposed symbols at the current \State position $j$ is $2$} 
\State  symbol $j$ of $f_x$ $\leftarrow$ the other  symbol at \State position $j$ of the superposed symbols
            \EndIf
        \EndFor
        \For{each end-device $y$ }
        \For{each symbol at position $j$ }
        \If{ $y \neq x$ and all symbols $j$ of $f_x$  $ \neq $ *  \State and the symbol $j$ of $f_y = $ * \State}  \State symbol $j$ of $f_y$ $\leftarrow$ the remaining \State symbol $j$ of the superposed  symbols 
            \EndIf
        \EndFor
      \EndFor
\EndFor
\EndWhile       
 \end{algorithmic}
    \end{minipage}\end{algorithm}
    
\subsection{Guessing the frame}
\label{subsection:GuessingFrame}
In this subsection, we show that the choice of symbols by the gateway has an impact on the number of bitmaps transmissions needed for each end-device. We refer to Fig.~\ref{FigSuperposition} to give an example of our proposed algorithm which is described by Algorithm \ref{alg:decoding}.

\textit{\underline{Step 1:}} The gateway sends a frame with the following arbitrary set of symbols $f_{G1} = (64, 0, 32)$. ED1 replies with the bitmap ${b_1^1}  = (1, 0, 1)$, ED2 replies with ${b_1^2} = (0, 1, 1)$, and ED3 with ${b_1^3} = (0, 0, 1)$. The current data frame of ED1 corresponds to $f_{1} = (64, *, 32)$,  the current data frame of ED2 corresponds to $f_{2} = (96, 0, 32)$, and the current data frame of ED3 corresponds to $f_{3} = (96, *, 32)$. 
 
 \textit{\underline{Step 2:}} Since some of the frames of the end-devices still contain missing symbols that cannot be deduced by elimination, the gateway sends another frame $f_{G2} = (96, 0, 32)$. ED1 replies with ${b_2^1} = (0, 0, 1)$, and ED3 with ${b_2^3} = (1, 0, 1)$. The updated frames of ED1 and ED3 remain the same as in \textit{Step 1}, (i.e $f_{1} = (64, *, 32)$, $f_{2} = (96, 0, 32)$ and $f_{3} = (96, *, 32)$). ED2 did not reply since its frame was decoded in \textit{Step 1}.

\textit{\underline{Step 3:}} Since some of the frames of the end-devices still contain missing symbols that cannot be deduced by elimination, the gateway sends another frame $f_{G3} = (96, 32, 32)$. ED1 replies with ${b_3^1} = (0, 1, 1)$, and ED3 with ${b_3^3} = (1, 0, 1)$. Now the updated frame of ED1 is $f_{1} = (64, 32, 32)$, and the updated frame of ED3 is $f_{3} = (96, 64, 32)$. Note that the second symbol of $f_{3}$ is decoded by deduction. 

In this example, the average number of transmissions for each end-device is 2.33 bitmap transmissions.

Another example to decode the aforementioned colliding frames is the following:

\textit{\underline{Step $1^{'}$:}} same as \textit{Step 1}

\textit{\underline{Step $2^{'}$:}} The gateway sends $f^{'}_{G2} = (96, 32, 32)$. ED1 replies with ${b_2^1} = (0, 1, 1)$, and ED3 with ${b_2^3} = (1, 0, 1)$. So the updated frames of ED1 and ED3 become $f_{1} = (64, 32, 32)$, and $f_{3} = (96, 64, 32)$ respectively. 

Here, the average number of bitmap transmissions for each end-device is 1.66 transmissions. 

Therefore, the choice of the symbols by the gateway may impact the number of needed bitmap transmissions for each end-device. 
Hence, we propose a random selection of symbols that are not already sent by the gateway.

\section{Simulations and results}
 In this section, we study and evaluate the performance of our proposed algorithms in terms of delay of successful decoding of colliding signals, energy consumption, and throughput.
 
 \subsection{Parameter settings}
 Simulations are carried out using our own simulator developed in Java. We model a network with a single gateway, and $x$ end-devices. We assume that all end-devices transmit with the same SF and on the same channel, and that their signals are received with almost the same strength at the gateway, i.e., no capture effect occurs. In our proposed algorithm, we assume that the time is divided into slots and that transmissions on the same slot are fully synchronized. We use both SF7 and SF12, as these two SFs have the highest and lowest data rates, respectively. The frame length is set to 30 bytes. We assume that collisions did not occur during the transmission of the bitmaps. Furthermore, we assume that the network is not saturated \footnote{We chose the non-saturated case in order to better analyze the network performance. The saturated case will be handled in an extension of this paper.}. 
We used $nmaxslots = 1000$, the maximum number of EDs $x = 8$, and the $gap = 30~nanoseconds$. Simulation results are obtained by averaging over one thousand samples.
 
  \subsection{Number of bitmap transmissions in our protocol vs number of frame retransmissions in LoRaWAN}
\begin{figure}[h]
  \centering
   \begin{tabular}{@{}c@{}}
    \label{fig:retransmissionsLoRa}\includegraphics[width=.8\linewidth,height=133pt]{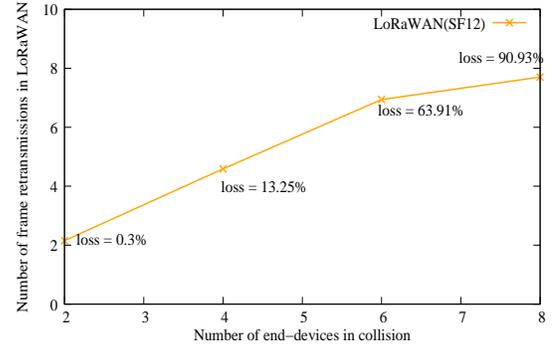} \\[\abovecaptionskip]
    \small (a) Number of frame retransmissions in LoRaWAN.
  \end{tabular}
  \vspace{\floatsep}
   \begin{tabular}{@{}c@{}}
   \label{fig:necessaryBitMaps} \includegraphics[width=.8\linewidth,height=133pt]{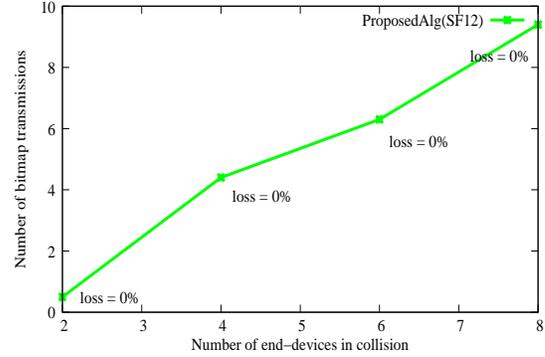} \\[\abovecaptionskip]
    \small (b) Number of bitmap transmissions in our algorithm.
  \end{tabular}
  \caption{Average number of transmissions per end-device in both LoRaWAN and our proposed algorithm.}
  \label{FigTransm}
\end{figure}

Figure~\ref{FigTransm}.a) shows the number of retransmissions attempts in LoRaWAN with the percentage of frame loss. Figure~\ref{FigTransm}.b) shows the number of necessary bitmaps for our proposed algorithm. Simulations are run using SF12. It is known that the number of collisions increases with the number of end-devices. This increase leads to an increase in the number of frame retransmissions in LoRaWAN, and to an increase in the number of necessary bitmaps needed to decode the colliding frames in our algorithm. The maximum number of retransmissions in LoRaWAN is set to 8 attempts by default \cite{alliance2015lorawan}. Hence, after 8 retransmissions attempts in LoRaWAN, a loss might arise for the colliding frames. For example, we observe in Fig.~\ref{FigTransm}.a), that for 8 colliding end-devices, we have almost 7.5 retransmissions per end-device with 90.93\% of frame losses, while in Fig.~\ref{FigTransm}.b), we have almost 9.5 bitmap transmissions per end-device without frame losses. Compared to \cite{rachkidy2018decoding}, in the case of two synchronized signals and when the two frames of the two EDs collide, each end-device has in average 0.5 frame retransmissions. In LoRaWAN, each end-device has in average 2.15 frame retransmissions with 0.3\% frame losses. Finally, in our proposed algorithm, each end-device has in average 0.5 bitmap transmissions (which is almost equivalent to 0.04 frame retransmissions, therefore twelve times better than \cite{rachkidy2018decoding}). Furthermore, the paper \cite{rachkidy2018decoding} does not handle the case of three or more colliding EDs. 

 


 \subsection{Average delay}
\begin{figure}[h]
   \includegraphics[scale=0.7]{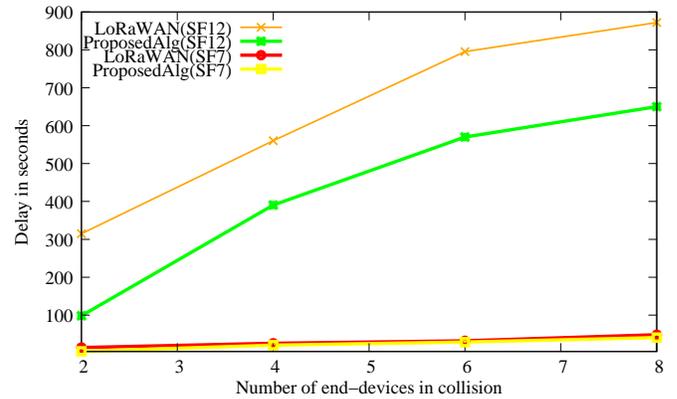}
  \caption{Delay of superposed signals.}
  \label{FigDelaySF12}
\end{figure}

Figure~\ref{FigDelaySF12} shows the evolution of the delay for the correct decoding of a frame in LoRaWAN and in our algorithm. 
The delay specifies the difference between the full decoding time of the frame by the gateway and the first time it is sent by the ED.
We notice that the delay in LoRaWAN is greater than the delay in our algorithm. This is due to the transmission of short bitmaps in our algorithm instead of the retransmission of the whole frame in LoRaWAN. Since the size of a bitmap is smaller than that of a whole frame, the transmission time of a bitmap is much smaller than the transmission time of a whole frame, which leads to decrease the time needed to decode the colliding frame in our algorithm compared to LoRaWAN. For example, for 4 colliding end-devices with SF12, we observe a decrease in the delay of 30\% in our algorithm compared to LoRaWAN. And for 4 colliding end-devices with SF7, we observe a decrease in the delay of 20\% in our algorithm compared to LoRaWAN. Moreover, it is obvious that the delay for the correct decoding of a frame under SF12, is greater than that under SF7. This is because SF7 has the shortest time on air (i.e. the shortest frame time duration), while SF12 has the longest time on air.

 \subsection{Average energy consumption}
\begin{figure}[h]
   \includegraphics[scale=0.7]{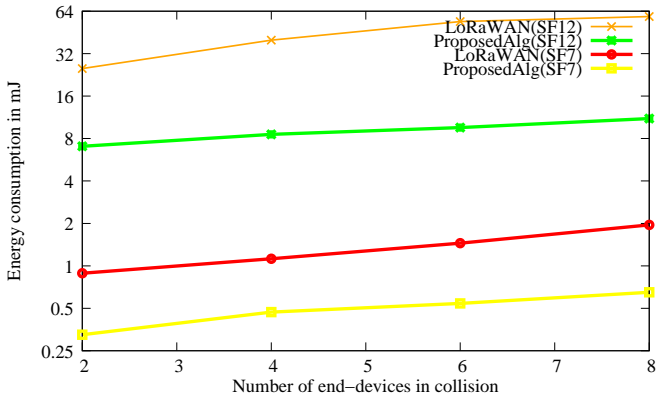}
  \caption{Energy consumption of superposed signals.}
  \label{FigECSF12}
\end{figure}

Figure~\ref{FigECSF12} shows the energy consumption calculated for an end-device after the full decoding of its frame by the gateway in both LoRaWAN and our proposed algorithm. It presents the consumed energy per useful bit as a function of the number of colliding signals at two spreading factors (SF7 and SF12). It is observed that the consumed energy increases with the increase of the number of colliding signals, and also with the increase of SF. As known, the greater value of SF, the more time is taken to send a frame, so the more consumed energy is needed to transmit data. We refer to the following equation to compute the consumed energy per useful bit: $E_{bit} = (\frac{P_{cons}(P_{tr})*(N_{Payload}+N_{P}+ 4.25)*2^{SF}}{8*PL*BW})$~\cite{bouguera2018energy}, where $P_{cons}(P_{tr})$ is the total consumed power that depends on transmission power. We set $P_{tr}$ to 14 dBm. $PL$ is the payload size set to 30 bytes. $N_{Payload}$ is the number of symbols used to transmit the payload and it is set to 30 for SF12 and 45 for SF7. $N_{P}$ is the preamble length set to 10. $BW$ is the bandwidth set to 125000 Hz. We observe that the energy consumption for an end-device in LoRaWAN is much greater than that for an end-device in our algorithm. This is due to the transmission of bitmaps in our algorithm instead of the transmission of whole frames in LoRaWAN, which leads to decrease the energy consumption of the colliding frame. We observe with SF7 a decrease of almost 65\% in the energy consumption in our algorithm in comparison with LoRaWAN, and we observe with SF12 a decrease of almost 78\% in the energy consumption in our algorithm in comparison with LoRaWAN. 

 \subsection{Average throughput}
\begin{figure}[h]
   \includegraphics[scale=0.7]{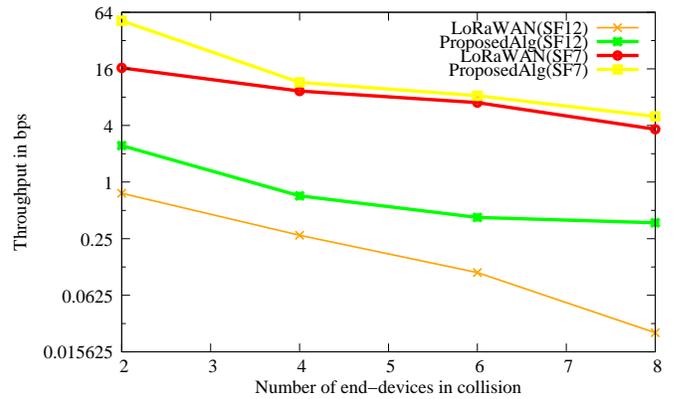}
  \caption{Throughput after a collision.}
  \label{FigThroughputSF12}
\end{figure}

Figure~\ref{FigThroughputSF12} shows the evolution of the throughput computed for an end-device at the gateway side in both LoRaWAN and our proposed algorithm. The throughput is already small in LoRaWAN, and it decreases further with the number of end-devices and collisions. In Fig.~\ref{FigThroughputSF12}, we observe that the throughput for an end-device in LoRaWAN is smaller than that for an end-device in our algorithm. This is due to the delay in LoRaWAN which is greater than the delay in our algorithm, and which leads to decrease LoRaWAN throughput compared to our algorithm. In addition, we have frame losses in LoRaWAN, which increase with the number of end-devices as already shown in Fig.~\ref{FigTransm}.a). This increase leads to decrease further the throughput in LoRaWAN compared to our algorithm where no frame losses are present. For instance, for 8 colliding end-devices with SF12, we observe a gain of almost 95\% in our algorithm compared to LoRaWAN. For 8 colliding end-devices with SF7, we observe a gain of almost 27\% in our algorithm compared to LoRaWAN.

 \section{Conclusion}
Collision is a factor that negatively impacts LoRaWAN throughput, which is already very limited (between 250 and 11000 bps). In this paper, we propose a collision resolution algorithm that enables to decode colliding frames in LoRa. Our algorithm focuses on the case where end-devices are fully synchronized. The proposed algorithm relies on retransmitting bitmaps in reply to guesses from the gateway instead of the whole frame. Based on our simulation results, we show that the proposed algorithm is able to improve the throughput, by decoding the frames in collision. This algorithm is also able to reduce the energy consumption of the end-devices, and to decrease the delay needed to decode the frames. These results contributed to the development of a new MAC protocol based on LoRaWAN, relying on the proposed collision resolution algorithm, and surpassing LoRaWAN.

\bibliographystyle{IEEEtran}
\bibliography{biblio}

\begin{thebibliography}{10}
\providecommand{\url}[1]{#1}
\csname url@samestyle\endcsname
\providecommand{\newblock}{\relax}
\providecommand{\bibinfo}[2]{#2}
\providecommand{\BIBentrySTDinterwordspacing}{\spaceskip=0pt\relax}
\providecommand{\BIBentryALTinterwordstretchfactor}{4}
\providecommand{\BIBentryALTinterwordspacing}{\spaceskip=\fontdimen2\font plus
\BIBentryALTinterwordstretchfactor\fontdimen3\font minus
  \fontdimen4\font\relax}
\providecommand{\BIBforeignlanguage}[2]{{%
\expandafter\ifx\csname l@#1\endcsname\relax
\typeout{** WARNING: IEEEtran.bst: No hyphenation pattern has been}%
\typeout{** loaded for the language `#1'. Using the pattern for}%
\typeout{** the default language instead.}%
\else
\language=\csname l@#1\endcsname
\fi
#2}}
\providecommand{\BIBdecl}{\relax}
\BIBdecl

\bibitem{talari2017review}
S.~Talari, M.~Shafie-khah, P.~Siano, V.~Loia, A.~Tommasetti, and J.~P.
  Catal{\~a}o, ``{A review of smart cities based on the Internet of Things
  concept},'' \emph{Energies}, vol.~10, no.~4, p. 421, 2017.

\bibitem{LoRaSemtech}
Semtech, ``{AN1200.22 LoRa™ Modulation Basics},''
  {https://www.semtech.com/uploads/documents/an1200.22.pdf}.

\bibitem{SigFox}
Sigfox, {http://www.sigfox.com.}

\bibitem{IngenuRPMA}
IngenuRPMA., {http://www.ingenu.com/technology/rpma/.}

\bibitem{Weightless}
W.~O. Standard., {http://www.weightless.org. Accessed:2015-11-07}.

\bibitem{alliance2015lorawan}
L.~Alliance, ``{LoRaWAN specification},'' \emph{LoRa Alliance}, 2015.

\bibitem{loraAlliance}
{https://www.lora-alliance.org/}.

\bibitem{rahmadhani2018lorawan}
A.~Rahmadhani and F.~Kuipers, ``{When LoRaWAN Frames Collide},'' 2018,
  https://fernandokuipers.nl/papers/WiNTECH2018.pdf.

\bibitem{ferre2017collision}
G.~Ferr{\'e}, ``Collision and packet loss analysis in a lorawan network,'' in
  \emph{Signal Processing Conference (EUSIPCO), 2017 25th European}.\hskip 1em
  plus 0.5em minus 0.4em\relax IEEE, 2017, pp. 2586--2590.

\bibitem{reynders2016chirp}
B.~Reynders and S.~Pollin, ``Chirp spread spectrum as a modulation technique
  for long range communication,'' in \emph{Communications and Vehicular
  Technologies (SCVT), 2016 Symposium on}.\hskip 1em plus 0.5em minus
  0.4em\relax IEEE, 2016, pp. 1--5.

\bibitem{petajajarvi2017performance}
J.~Pet{\"a}j{\"a}j{\"a}rvi, K.~Mikhaylov, M.~Pettissalo, J.~Janhunen, and
  J.~Iinatti, ``{Performance of a low-power wide-area network based on LoRa
  technology: Doppler robustness, scalability, and coverage},''
  \emph{International Journal of Distributed Sensor Networks}, vol.~13, no.~3,
  p. 1550147717699412, 2017.

\bibitem{liao2017multi}
C.-H. Liao, G.~Zhu, D.~Kuwabara, M.~Suzuki, and H.~Morikawa, ``{Multi-hop LoRa
  networks enabled by concurrent transmission},'' \emph{IEEE Access}, vol.~5,
  pp. 21\,430--21\,446, 2017.

\bibitem{eletreby2017empowering}
R.~Eletreby, D.~Zhang, S.~Kumar, and O.~Ya{\u{g}}an, ``{Empowering Low-Power
  Wide Area Networks in Urban Settings},'' in \emph{Proceedings of the
  Conference of the ACM Special Interest Group on Data Communication}.\hskip
  1em plus 0.5em minus 0.4em\relax ACM, 2017, pp. 309--321.

\bibitem{rachkidy2018decoding}
N.~El~Rachkidy, A.~Guitton, and M.~Kaneko, ``{Decoding Superposed LoRa
  Signals},'' \emph{IEEE LCN}, 2018.

\bibitem{bouguera2018energy}
T.~Bouguera, J.~Diouris, J.~Chaillout, R.~Jaouadi, and G.~Andrieux, ``{Energy
  Consumption Model for Sensor Nodes Based on LoRa and LoRaWAN.}''
  \emph{Sensors (Basel, Switzerland)}, vol.~18, no.~7, 2018.

\end{thebibliography}

\end{document}